\documentclass[%
 reprint,
superscriptaddress,
 amsmath,amssymb,
 aps,
pra,
prper,
]{revtex4-2}

\usepackage[table,xcdraw]{xcolor}

\usepackage{physics}
\usepackage{xfrac}
\usepackage{braket}
\usepackage{amsmath}
\usepackage{float} 

\usepackage{graphicx}

\usepackage{xcolor}   
\usepackage{hyperref} 

\definecolor{myblue}{rgb}{0, 0, 1}
\definecolor{mygreen}{rgb}{0, 1, 0}

\hypersetup{
    colorlinks=true,
    linkcolor=myblue,
    citecolor=myblue,
    urlcolor=myblue
}

\usepackage{dcolumn}
\usepackage{bm}

\usepackage{chemformula}

\begin{document}


\title{\ch{C60} fullerene as an on-demand single photon source at room temperature}

\author{Raul Lahoz Sanz}
 \email{rlahozsanz@icc.ub.edu}
\affiliation{%
Departament de Física Quàntica i Astrofísica, \\
Facultat de Física, Universitat de Barcelona (QCommsUB group)
}%
\affiliation{Institut de Ciències del Cosmos (ICCUB), Universitat de Barcelona (UB), c. Martí i Franqués, 1, 08028 Barcelona, Spain}

\author{Lidia Lozano Martín}
\affiliation{Institut de Ciències del Cosmos (ICCUB), Universitat de Barcelona (UB), c. Martí i Franqués, 1, 08028 Barcelona, Spain}
\affiliation{
Department of Applied Physics, Universitat de Barcelona, C/Martí i Franquès 1, 08028, Barcelona, Spain
}

\author{Adrià Brú i Cortés}
\affiliation{Departament d’Enginyeria Electrònica i Biomèdica, Universitat de Barcelona (UB),  c. Martí i Franqués, 1, 08028 Barcelona, Spain}

\affiliation{Institut de Ciències del Cosmos (ICCUB), Universitat de Barcelona (UB), c. Martí i Franqués, 1, 08028 Barcelona, Spain}

\author{Sergi Hernández}
\affiliation{
Departament d’Enginyeria Electrònica i Biomèdica, Universitat de Barcelona (UB),  c. Martí i Franqués, 1, 08028 Barcelona, Spain
}
\affiliation{
Institute of Nanoscience and Nanotechnology (IN2UB), Universitat de Barcelona (UB), 08028, Barcelona, Spain
}

\author{Martí Duocastella}
\affiliation{
Department of Applied Physics, Universitat de Barcelona, C/Martí i Franquès 1, 08028, Barcelona, Spain
}
\affiliation{
Institute of Nanoscience and Nanotechnology (IN2UB), Universitat de Barcelona (UB), 08028, Barcelona, Spain
}

\author{José M. Gómez-Cama}
\affiliation{Departament d’Enginyeria Electrònica i Biomèdica, Universitat de Barcelona (UB),  c. Martí i Franqués, 1, 08028 Barcelona, Spain}

\affiliation{Institut de Ciències del Cosmos (ICCUB), Universitat de Barcelona (UB), c. Martí i Franqués, 1, 08028 Barcelona, Spain}

\affiliation{Institut d'Estudis Espacials de Catalunya (IEEC), Edifici RDIT, Campus UPC, 08860 Castelldefels (Barcelona), Spain}

\author{Bruno Juliá-Díaz}
 \email{brunojulia@ub.edu}
\affiliation{%
Departament de Física Quàntica i Astrofísica, \\
Facultat de Física, Universitat de Barcelona (QCommsUB group)
}%
\affiliation{Institut de Ciències del Cosmos (ICCUB), Universitat de Barcelona (UB), c. Martí i Franqués, 1, 08028 Barcelona, Spain}

\begin{abstract}

Single photon sources are fundamental for applications in quantum computing, secure communication, and sensing, as they enable the generation of individual photons and ensure strict control over photon number statistics. However, current single photon sources can be limited by a lack of robustness, difficulty of integration into existing optical or electronic devices, and high cost. In this study, we present the use of off-the-shelf \ch{C60} fullerene molecules embedded in polystyrene as room-temperature reliable single-photon emitters. As our results demonstrate, these molecules exhibit on-demand single-photon emission, with short fluorescence lifetimes and, consequently, high emission rates. The wide availability and ease of preparation and manipulation of fullerenes as single photon sources can pave the way for the development of practical, economic and scalable quantum photonic technologies.

\end{abstract}

\maketitle

\section{Introduction}

Single photon sources (SPSs) have become a cornerstone in present day 
quantum technology applications in sensing~\cite{minns2024device}, 
quantum computing~\cite{couteau2023applications, franson2004quantum} and 
quantum communications. They are crucial for quantum 
key distribution (QKD)~\cite{bennett2014quantum, yang2024high}, where
encoding information in a single photon helps prevent potential photon 
number splitting attacks~\cite{lutkenhaus2002quantum}. The growing number of applications has driven research toward the development of optimal sources of individual photons. These should ideally be bright, stable, and easy to fabricate and operate. 

Brightness  - the ability to emit a high number of photons per unit time - is closely related to the radiative excitonic state's decay time, and ultimately dictates the speed at which information can be exchanged. Stability, on the other hand, can be compromised by the intermittency of the emission, referred to as blinking, or by a gradual decline of the emission over time, known as bleaching. A low production cost combined with room temperature operation is essential for the transition toward large-scale implementation. Added to the previous requirements is the need for on-demand single-photon generation, that is, the possibility to trigger emission using an external and controllable signal. To this end, the source must rely on discrete energy levels, as in quantum dots or atomic systems, where strong carrier confinement leads to 
quantized energy levels. It is also important to select the distribution of these levels. For instance, fast, nonradiative, Auger recombination can help suppress multi-exciton emission~\cite{xie2022nonblinking}, increasing the likelihood that only a single photon is emitted per excitation cycle. 

\begin{figure*}[t] 
    \includegraphics[width=1\textwidth]{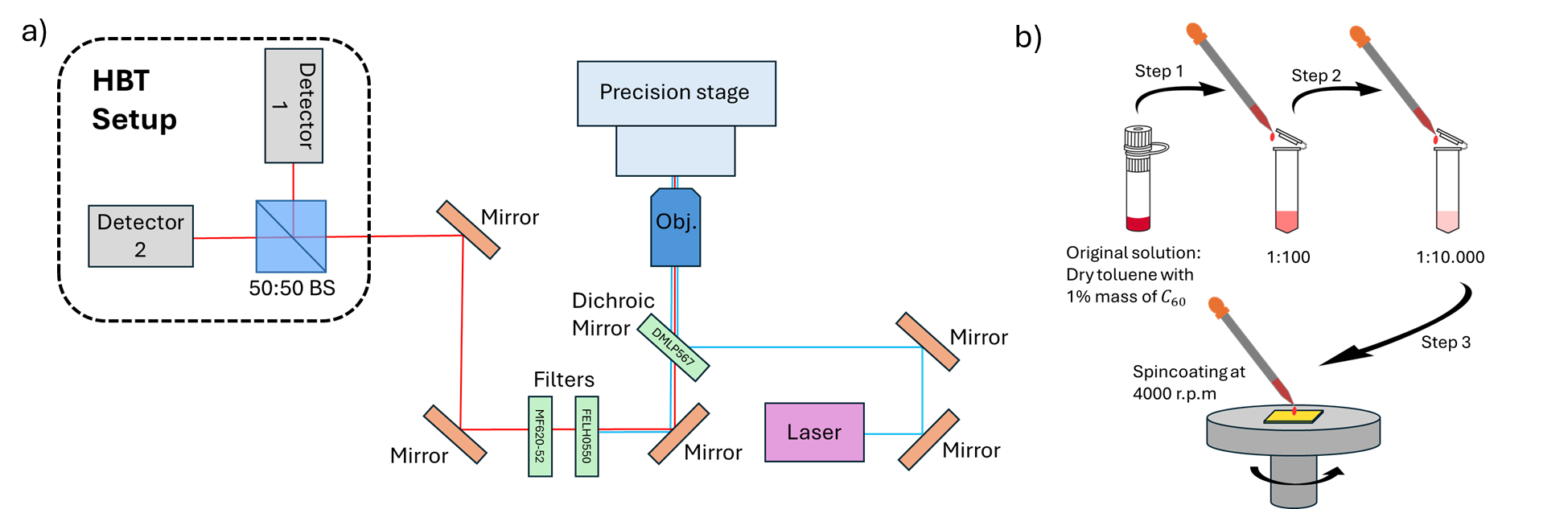}
    \caption{a) Scheme of setup used for the excitation and light 
    collection of an individual single photon source. The Hanbury Brown 
    and Twiss (HBT) setup used for measuring the second-order 
    autocorrelation function is also shown. The $50:50$ non-polarizing beam splitter ($50:50$ BS) effectively splits the light in two channels. 
    The first one (Ch 1) leads the light to the detector 1, the start 
    detector, while the second one (Ch 2) leads the light to the detector 2, 
    the stop detector. b) Schematic of the sample preparation process. Starting with a dry toluene sample in which 1$\%$ by mass of fullerene is dissolved. From this original solution, a serial dilution 
    is performed to obtain two additional samples with concentrations of 1:100 and 1:10,000, using dry toluene as the diluent. Finally, a drop is deposited onto a gold-coated substrate while it is spinning at 4000 rpm.}
    \label{fig:SetupScheme}
\end{figure*}

Recent technological progress has brought the concept of an ideal SPS closer to reality. However, the most commonly used sources - epitaxial quantum dots, colloidal quantum dots, and 
nitrogen-vacancy (NV) centers in nanodiamonds~\cite{abe2017dynamically, hirt2021sample} - still present significant trade-offs. Epitaxially grown quantum dots offer high 
brightness and excellent emission stability~\cite{limame2024high, holewa2022bright}, but their production cost is high and require cryogenic temperatures during operation~\cite{holewa2020thermal}. 
Colloidal quantum dots (CQDs) differ from epitaxial 
ones in that they exhibit strong quantum confinement, which keeps carriers 
trapped in discrete energy levels and preserves single-photon purity even at 
room temperature. They are inexpensive to produce, but typically exhibit 
average exciton lifetimes of around 20 ns when not coupled to optical 
cavities~\cite{ihara2019superior, lin2017electrically}. These sources suffer 
from pronounced emission intermittency~\cite{panev2001random} and relatively 
short operational lifetimes, often limited to just a few minutes, which 
ultimately restricts their brightness~\cite{qin2017photoluminescence}. An alternative 
that addresses the issue of emission intermittency is the use of NV-centers 
in nanodiamonds~\cite{kurtsiefer2000stable, gaebel2012size, rodiek2017experimental}. 
In these systems, blinking is almost completely suppressed, and in some cases, 
stable emission can be maintained for several hours. However, the brightness of these sources is typically low, as only a small 
fraction of the emitted photons can escape the nanodiamond and reach 
the far field. The broadband spectra of these sources can also be a drawback, 
especially for applications requiring monochromatic light or when coupling to resonant cavities is needed.

Alternatively, it is possible to use molecules as 
 sources of single-photon emission. Examples range from optical 
excitation in Oxazine 720~\cite{de1996single} or terrylene 
molecules~\cite{lounis2000single, treussart2002single} to electrical 
excitation in ZnPc molecules~\cite{zhang2017electrically}. Single molecules 
present several appealing characteristics for single-photon emission. Their 
large transition dipole moments and short spontaneous emission lifetimes 
contribute to high brightness and emission efficiency. However, single 
molecules also face notable limitations. Issues such as photobleaching and 
emission intermittency can hinder their long-term operation, while environmental 
sensitivity  introduces instability and reduces the efficiency of single 
photon emission~\cite{gaither2023organic}.

Here, we show how  \ch{C60} fullerene molecules can operate as novel, low-cost, and room-temperature
single-photon sources. In addition, their significantly shorter radiative lifetime compared to other SPSs, such as CQDs, enables on-demand high emission rates. 
To the best of our knowledge, the functionality of \ch{C60} fullerenes as single-photon sources (SPSs) has only been demonstrated by excitation via charge injection through quantum tunneling from the tip of a scanning tunneling microscope cantilever~\cite{merino2015exciton}. Our approach is more accessible and 
easier to implement, relying on the optical excitation of off-the-shelf \ch{C60} 
fullerene molecules embedded in polystyrene. We conduct an extensive 
study of their emission properties, including anti-bunching experiments, measurement of the exciton lifetimes and analysis of their 
blinking statistics, both under continuous as well as pulsed light excitation. 

\section{Methods}

To prepare our samples, we use dry toluene with 5$\%$ (w/w) polystyrene. 
In this solution, we dissolve 1$\%$ by mass of our fullerene sample 
(\textit{Sigma Aldrich, 379646}). Then, we begin a serial dilution, 
obtaining successive dilutions of 1:100 and 1:10000 from the initial 
solution, where the solvent is our dry toluene with 5$\%$ mass of 
polystyrene. This dilution process is carried out inside a glovebox 
with an $N_{2}$ atmosphere.

Next, we deposit a 5~$\mu$L drop onto a gold-coated silicon wafer, 
while the wafer rotates at 4000~rpm inside the spincoater. At the 
moment the drop is deposited, we allow it to spin 
for 1~min to ensure proper spreading of the drop.

Once the droplet has spread across the entire gold-coated surface, 
the rapid evaporation of the toluene leads to the formation of a polystyrene layer in which the fullerene molecules are embedded. 
Here, the polystyrene film formed during spin coating helps to 
preserve the sample during a long period of time~\cite{raino2019underestimated}. 

After sample preparation, we use the setup depicted in Fig.~\ref{fig:SetupScheme} to excite and collect light from an 
individual single photon source. For the optical excitation, we use 
a 405~nm laser (\textit{USB-Powered Laser Module, Flim Labs}), that 
can work both in the continuous and pulsed regime, with a pulse 
duration of 50~ps. The laser light is focused on the sample with 
an air-based Nikon objective (0.9 N.A., 100$\times$). This blue light 
excites our sample, and the emitted light from the source, 
upon de-excitation, is also collected by the objective and 
directed to a dichroic mirror (\textit{DMLP567, ThorLabs}) with 
a cut-off wavelength of 567~nm. Light transmitted through the 
dichroic mirror, i.e. wavelengths above 567~nm, also pass 
through a long-pass filter (\textit{FELH0550, ThorLabs}) and a 
band-pass filter (\textit{MF620-52, ThorLabs}). This ensures 
that only light with a wavelength around 620~nm is detected, 
while also preventing any residual blue laser light from 
contaminating our measurements. Finally, light is sent to a 
Hanbury Brown and Twiss (HBT) setup for measuring the 
second-order autocorrelation function, see Fig.~\ref{fig:AutocorrelationFunctions} a).

The HBT setup consists of a 50:50 non-polarizing beam splitter that 
splits each photon into a superposition of being on each arm of 
the HBT. Each arm guides the light into a fiber coupler. After coupling into a multimode fiber, the photons are directed 
to two different channels of our single photon detector 
(\textit{SPCM-AQ4C, Excelitas}) based on silicon
avalanche photodiodes. The electronic pulses produced by the detector upon photon detection are sent to a time-correlated 
single photon counting (TCSPC) module (\textit{TT-Ultra, Swabian Instruments}). This device records the exact arrival time of 
each event at each channel, as well as the synchronization pulses 
from the pulsed laser when it is working in the pulsed 
wavelength (PW) mode. This allows us to compute the second-order autocorrelation function by plotting all the different time 
delays between events detected in different channels of the HBT. 
Since our TCSPC module has a precision of about 50~ps and our single-photon detectors have a precision of around 600~ns, the main source of error in measuring the arrival time of each photon comes from the detectors. 
In contrast, the measurement of the synchronization signal 
for each laser pulse is only affected by the TCSPC module’s uncertainty, which is 50~ps.

\section{Results and discussion}

First, we characterize the emission properties of the \ch{C60} molecules using a highly concentrated sample (1:100). The emission spectrum of the molecules shows a band 
centered at 2~eV (620~nm), as seen in the photoluminescence spectra reported in Fig.~\ref{fig:photoactivation}. Interestingly, the location of this peak is not consistent with 
self-trapped polaron emission~\cite{Matus1992}. A possible explanation of the observed spectrum is photo-induced emission, in which 
light emission is enhanced or altered upon exposure 
to ultraviolet or visible light (see Supporting Information for further details). As shown in Fig.~\ref{fig:photoactivation}, a progressive enhancement of emission 
from our \ch{C60} molecules is observed 
upon excitation with 405~nm laser light, consistent with the phenomenon of photo-induced emission. This process can be explained by the photooxidation of \ch{C60} when using photons with energies within the 
absorption band of the molecule. Thus, the molecules 
undergo oxidation by reacting with molecular oxygen, resulting in the formation of oxidized fullerene species 
with lower symmetry, such as \ch{C60O_{n}}. The lower symmetry 
enhances the HOMO-LUMO (highest occupied molecular orbital - lowest unoccupied molecular orbital) transition. This transition was originally forbidden in the highly 
symmetric pristine \ch{C60}, thereby increasing the 
photoluminescence emission. This 
oxidative modification leads to a permanent increase in 
fluorescence intensity and a spectral blue shift observed 
during the irradiation process~\cite{zhang1996photoluminescence}. 

Although all evidence points to the emitting species being oxidized fullerene molecules, \ch{C60O_{n}}, the exact number of oxygen atoms attached to each fullerene 
molecule cannot be determined with our 
current experimental setup. Still, the Raman spectrum of one of our samples 
features a significant contribution attributed to the oxidation of 
fullerene molecules, see Fig.~\ref{fig:raman}. This indicates that 
these oxidized species are prevalent in our samples. Therefore, from 
this point on, whenever we refer to a fullerene molecule behaving as 
a single-photon source, we will be referring to an 
oxidized fullerene molecule of the type \ch{C60O_{n}}.  
The exact number of oxygen atoms attached to each 
molecule in a given experiment is difficult to determine 
with our experimental system.

\begin{figure*}[t] 
   
    \includegraphics[width=1\textwidth]{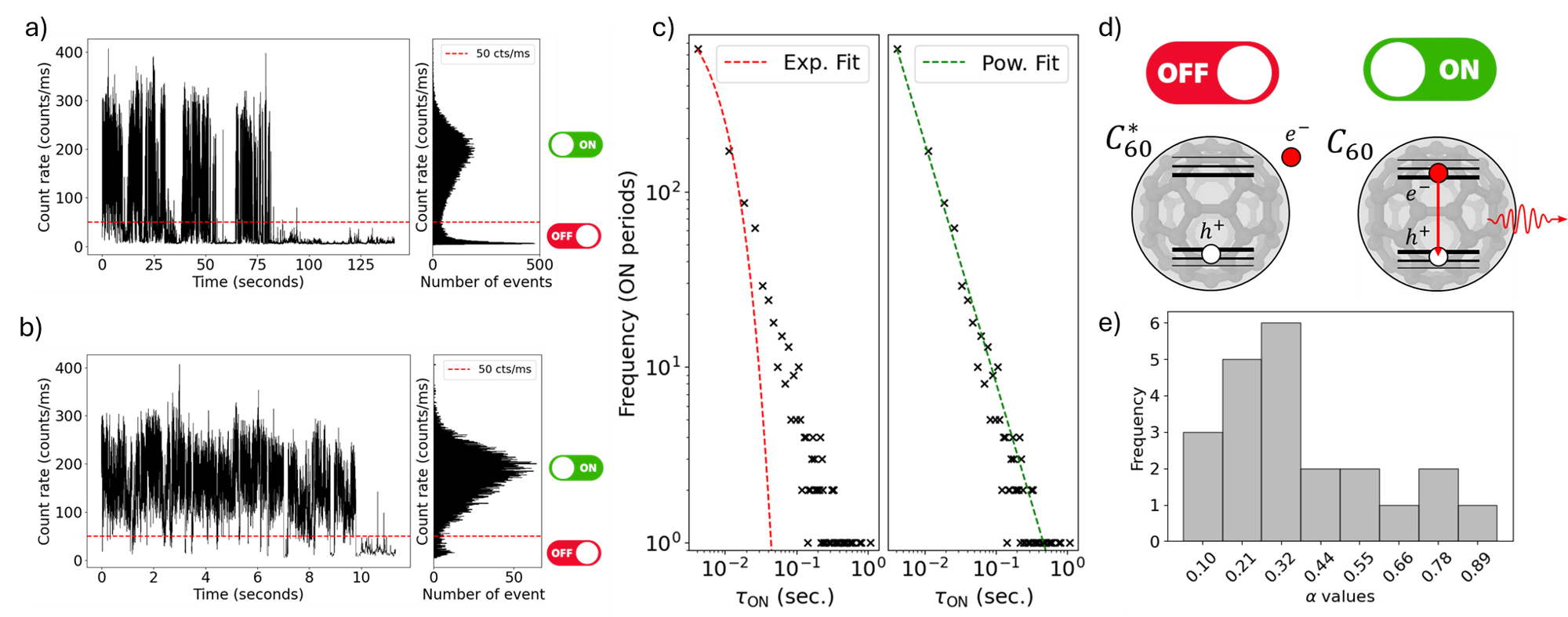}
    \caption{Blinking behavior of our sources. a) PL intensity of the emission during all the measurement. b) Detail of the first 12 seconds of the PL intensity measurement. In both graphs, the ON and OFF switches represent the periods when our source is emitting or not, respectively, while the threshold separating these two behaviors is set at 50 counts/ms. c) Graph illustrating the fits to an exponential curve (left) and a power-law curve (right) for the duration of the different periods in which our source remained in the ON state ($\tau_{\rm{ON}}$). d) Scheme illustrating the state of our sources during ON and OFF periods. e) Histogram representing the different values obtained of $| \alpha |$ in the power-law fitting in different \ch{C60} molecules.}
    \label{fig:Blinking}
\end{figure*}

After demonstrating light emission from \ch{C60} molecules, we next investigate their behavior as single photon emitters. To this end, we characterize the temporal emission of a highly diluted sample (1:10,000), i.e. the molecules are sparse enough to be analyzed individually. 
Fig.~\ref{fig:Blinking} a) and b) show a characteristic PL emission 
trace from a single \ch{C60} molecule under CW 
excitation. Notably, our source features blinking, i.e. fluctuations between two 
distinct emission rates, which is a strong  
indication that we are exciting a single emitter.
Blinking can be attributed 
to Auger ionization. In this process, the fullerene molecule becomes temporarily ionized when either an electron or a hole is expelled from the interior, see Fig.~\ref{fig:Blinking} d). During this period, the source does not exhibit radiative emission~\cite{efros1997random, kuno2000nonexponential}.

Further analysis of the emission statistics of individual \ch{C60} molecules can reveal more details about their photophysical properties. The distribution of counts features a marked bimodal distribution, see Fig.~\ref{fig:Blinking} a) and b). This can 
be attributed to the constant switching between two different states, i.e. ON and OFF. 
In our analysis, we fixed a threshold of 50 counts per millisecond to separate the ON and OFF states. The average count rate detected for the ON state is around 200 counts/ms, while for the OFF state is around 20 counts/ms. It is worth mentioning that the dark count rate from our detectors is 1 count/ms. 
The difference between the count rate for the OFF state and the expected count rate due to the dark counts may be due to the fluorescence of the sample and the background noise of the laboratory.
The distribution of the duration of the ON periods, $\tau_{\rm ON}$, can be 
related to the three- or multi-state nature of the photon 
emission, ~\cite{efros1997random, kuno2000nonexponential}. 
The three-level description predicts an exponential decay of the distribution~\cite{efros1997random}. 
In our case, our data does not follow an exponential decay. Instead, we observe a clear power-law behavior $\propto  \tau_{\rm ON}^{- (1 + \alpha )}$, see Fig.~\ref{fig:Blinking} c), indicating that the emission of individual photons requires a more complex model~\cite{kuno2000nonexponential}. Note that the particular molecule analyzed in the figure exhibits a value of $\alpha$  = 0.369~$\pm$~0.011. 
Repeating the measurement, a distribution of $\alpha$ values is obtained, as shown in Fig.~\ref{fig:Blinking} e). All values fall within the interval $0 < \alpha < 1$~\cite{kuno2001off}. 
Importantly, this type of intermittency behavior in the emission of single-photon sources has already been previously observed in different types of emitters, such as colloidal quantum dots~\cite{rabouw2019microsecond, kuno2001off} and NV-centers in nanodiamonds~\cite{gu2013super}.

\begin{figure*}[t] 
    \includegraphics[width=1\textwidth]{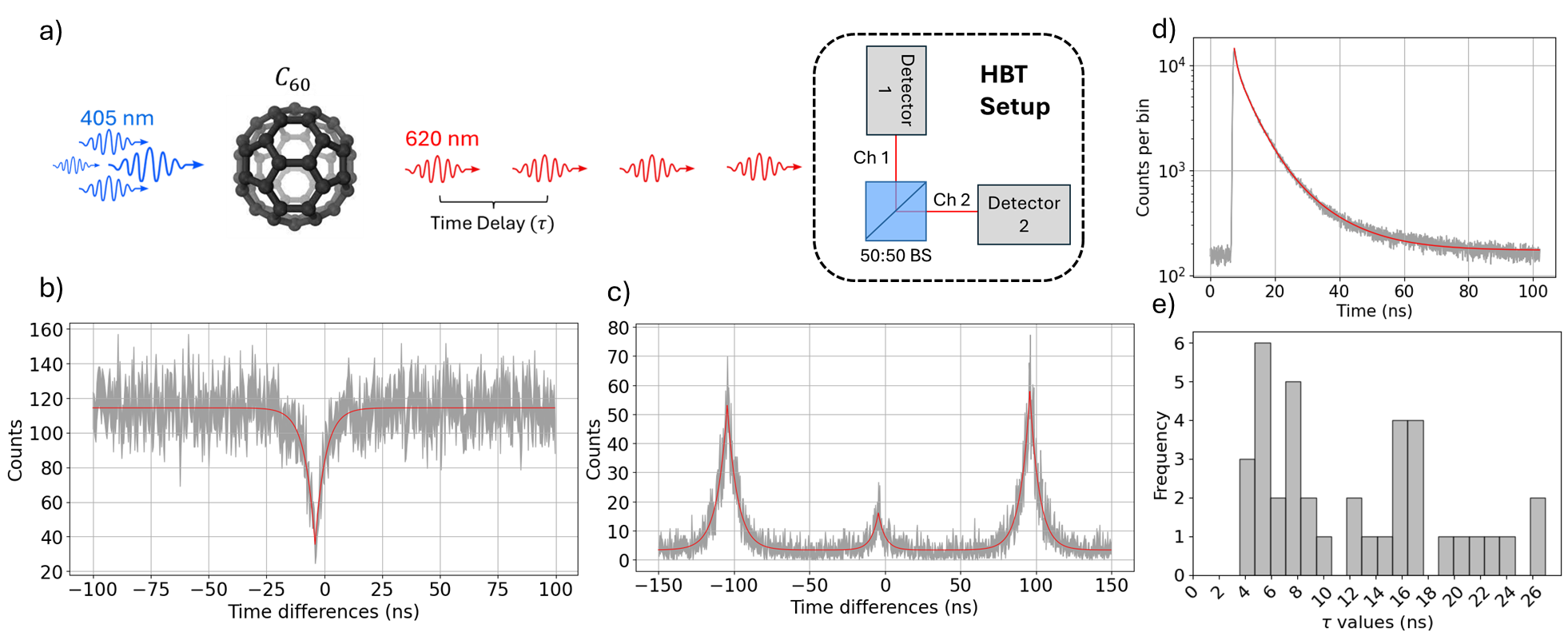}
    \caption{ a) Scheme of the second-order autocorrelation 
    measurement. We illustrate the process involving the excitation of a 
    single \ch{C60} molecule with 405~nm laser light and the subsequent 
    antibunched emission of light around 620~nm is directed towards the HBT setup 
    for analyzing its emission statistics.  b), c) second-order autocorrelation 
    functions for the same SPS under CW excitation and PW excitation, respectively. 
    In both graphs, the bin width is 500~ps. d) Lifetime histogram of the same 
    SPS used for obtaining the CW and PW second-order autocorrelation measurements. 
    e) Histogram representing the dispersion of lifetimes values obtained 
    for different single photon sources. For the graphs b), c) and d), the values 
    shown on the 'y' axis correspond to the actual values along with 
    their corresponding error bars.}
    \label{fig:AutocorrelationFunctions}
\end{figure*}

The blinking behavior of our sources indicates the single 
photon emitting nature of our sources. For a more stringent 
demonstration of the operation of our source as a single 
photon emitter, we characterize the antibunching behaviour of the 
emitted light. We rely on the second-order autocorrelation function. It is calculated by plotting a histogram of the time delays between photon detection events recorded in different detectors of the HBT setup depicted in Fig.~\ref{fig:SetupScheme} a), where the time of arrival of each photon is recorded by the TCSPC. 
The fact that the value of the normalized second-order autocorrelation function around zero time delays is below 0.5, ($g^{(2)}(\tau ~ 0) < 0.5$) indicates that the source under investigation behaves as a single-photon emitter~\cite{grunwald2019effective}.

To obtain the second-order autocorrelation graph under continuous wavelength 
(CW) excitation, we use 
the 405~nm laser with an intensity of 0.891~kW/cm$^2$, measured at the 
entrance of the objective, as shown in Fig.~\ref{fig:AutocorrelationFunctions} b). The experimental points are correctly fitted by 
the function
\begin{equation}
    g^{(2)}_{\rm{CW}}(\tau) = a \left( 1 - b\cdot e^{\frac{-|\tau - \tau_0|}{\tau_X}} \right),
\label{g2_CW}
\end{equation}
where $\tau$ is the delay time between events. $\tau_0$ is the time offset between 
the two arms of the HBT setup and $\tau_X$ is the lifetime of the exciton. Once 
we have the value of all these variables, we obtain the normalized second-order 
autocorrelation function as,
\begin{equation}
    g^{(2)}_{\rm{norm}}(\tau) = \frac{1}{a} \cdot g^{(2)}(\tau)\,.
\label{g2_CW_norm}
\end{equation}
The function is normalized such that $ g^{(2)}_{\rm{norm}}\to 1$ for time delays 
much larger than the lifetime of the exciton. The value obtained for $\tau_X$ 
in Fig.~\ref{fig:AutocorrelationFunctions} b) is 4.700~$\pm$~0.692~ns, while 
the value of the normalized $g^{(2)}$ function at $\tau = \tau_0$ is 0.304~$\pm$~0.024~ns. Notably, the value of the normalized second-order autocorrelation function around time delays equal to zero is below 0.5, which indicates that the source under investigation behaves as a single-photon emitter.

Next, we repeat the experiments under pulsed illumination. Working under pulsed excitation is of particular interest as it allows us to achieve on-demand photon emission. In this mode, after each excitation pulse, the source emits one photon (or none, in case the excitation was unsuccessful or the source was ionized). Using a laser with pulse durations much shorter than the characteristic decay lifetime of our sources is crucial to avoid multiple excitations within a single laser pulse.
Thus, with the same \ch{C60} molecule and changing to PW excitation with a repetition rate 
of 10~MHz and an intensity of 0.114 kW/$cm^2$, we can obtain the second-order autocorrelation 
graph shown in Fig.~\ref{fig:AutocorrelationFunctions} c). In this case, the 
experimental points are fitted with, 
\begin{equation}
    g^{(2)}_{\rm PW}(\tau) = a + b_0 \cdot e^{\frac{-|\tau - \tau_{0}|}{\tau_X}} + \sum_{n \neq 0} b_n \cdot e^{\frac{-|\tau - \tau_{0} - n\cdot T|}{\tau_X}} \cdot \left( 1 - e^{\frac{-|\tau - \tau_{0}|}{\tau_X}}\right)\,.
\label{g2_PW}
\end{equation}
Where $b_i$ is the height in number of counts of each of the different 
peaks and $T$ is the time interval between pulses~\cite{ihara2019superior}. We use 
the mean value of the height in counts of all the different peaks surrounding 
the peak at $\tau = \tau_0$ to obtain the normalized second-order autocorrelation function
\begin{equation}
    g^{(2)}_{\rm{norm}}(\tau) = \frac{1}{ \bar{b}_{n \neq 0}} \cdot g^{(2)}(\tau)\,.
\label{g2_PW_norm}
\end{equation}
Here $\bar{b}_{n \neq 0}$ is the mean value of all the $b_n$ with $n\neq0$. The value 
of the normalized function at $\tau = \tau_0$ is equal to 0.308~$\pm$~0.024~ns, 
while the value for the lifetime of the exciton is 4.537~$\pm$~0.655~ns. As in the case of CW illumination, the minimum value of the second-order autocorrelation function around time delays equal to zero is below 0.5, indicating the single-photon emission nature of the analyzed source.

Finally, we measure the decay lifetime of \ch{C60} molecules, which is directly linked to their emission rate and, consequently, their potential use in high-speed quantum communications. To this end, we couple all the 
light emitted by our single-photon sources into a detector 
and record the arrival times of the detected counts. The 
corresponding plot of the time differences between each photon arrival and the 
preceding synchronization pulse of the pulsed laser is shown 
in Fig.~\ref{fig:AutocorrelationFunctions}~d). 
The data is well fitted by a multi-exponential function~\cite{andrewawang2013cadmium}
\begin{equation}
    f(\tau) = A + \sum_{i = 1}^{3} B_i \cdot e^{-\frac{\tau - \tau_0}{\tau_i}}\,.
\label{eq:multiexponential}
\end{equation}
The fitted values of the different decay lifetimes are $\tau_1$ = 0.833~$\pm$~ 0.015~ns, $\tau_2$ = 4.959~$\pm$~0.095~ns and $\tau_3$ = 13.135~ $\pm$~0.450~ns. 
We can interpret the first decay time as corresponding to 
the biexciton, the second to the exciton, and the third to 
a long-lived state. This already suggests that the emission level scheme 
of our system involves a more complex dynamics 
than what a simple two-level model would imply. The averaged 
decay lifetime, can be calculated as
\begin{equation}
    \tau_{\rm{avg.}} = \frac{\sum_{i = 1}^{3} B_i \cdot \tau_i}{\sum_{i = 1}^{3} B_i}\,,
    \label{eq:tau_avg}
\end{equation}
which leads us to a value of $\tau_{avg.}$ = 4.516~$\pm$~0.079~ns.

Interestingly, we find different regimes in the measured 
exciton lifetime for different measurements: in some cases, we observe that they are on 
the order of around 4~ns, while in others, it extends to 
around 20~ns, as shown in Fig.~\ref{fig:AutocorrelationFunctions} e). This 
variability in radiative decay times suggests 
the existence of different local environments or emissions 
pathways that affect the emission dynamics. Furthermore, decay 
lifetimes of around 4~ns are much shorter than those 
typically observed in other single-photon sources, such as 
colloidal quantum dots or NV-centers under similar conditions 
and without any cavity coupling~\cite{ihara2019superior, messin2001bunching, lin2017electrically, vonk2021biexciton, kurtsiefer2000stable}. Such short emission lifetimes result in higher emission rates, making these sources brighter, as more photons are emitted within a fixed time interval.

\section{Conclusions}

Fullerene \ch{C60} molecules can function as 
single-photon emitters at room temperature. Importantly, their emission can be triggered on demand by using pulsed laser excitation. As our results demonstrate, blinking effects are significant in this type of sources, characterized with short and bright periods of emission. A further analysis of the distribution of the durations of the ON and OFF periods revealed that \ch{C60} molecules cannot be modeled with a three-level system. Despite the blinking, the source maintains a high degree of single-photon purity during the ON periods, making it a promising candidate for applications in quantum communication and quantum information processing.

Regarding the mechanisms that enable emission in these particles, our results suggest that 405~nm wavelength excitation plays a key role, particularly triggering the photo-assisted emission phenomena we observed, especially in highly concentrated samples. This process, caused by the photo-oxidation of fullerene molecules, increases their emissivity by breaking molecular symmetry through the addition of oxygen atoms to the structure. This change allows transitions that were previously forbidden.

While the exact nature of single-photon emission at the molecular level remains uncertain, we hypothesize that oxidation processes occurring either in individual molecules or within clusters may be responsible for this behavior. Such a mechanism could also help explain the significant variability observed in the exciton lifetimes across different measurements.

Overall, the wide availability of \ch{C60}, along with its low production cost and ease of preparation, marks a significant step towards the practical implementation of these molecules as single-photon sources in quantum technologies.

\section*{Acknowledgements}
The authors thank Prof. Fei Ding and his research group in Hannover, Germany, 
for their discussions on the design of the setup. This study was supported by 
MCIN with funding from European Union NextGenerationEU(PRTR-C17.I1) and by 
Generalitat de Catalunya. We acknowledge funding from Grants 
PID2023-147475NB-I00 and CEX2024-001451-M financed by MCIN/AEI/10.13039/501100011033 
and Grants 2021SGR01095,  and 2021SGR01108 by Generalitat de Catalunya. 
This project has received funding from the European Union’s Digital 
Europe Programme under grant agreement no. 101084035.

\bibliographystyle{apsrev4-2}
\bibliography{biblio}

\begin{widetext}

\appendix

\clearpage

\begin{center}
    {\large \textbf{Supporting Information: \ch{C60} fullerene as an on-demand single photon source at room temperature}}
\end{center}

\section{Photoactivation of \ch{C60} molecules embedded in polystyrene}
\label{sec:Photoactivation}

As reported in Zhang {\it et al.}~\cite{zhang1996photoluminescence}, where \ch{C60} molecules embedded in polystyrene exhibit notable changes in their photoluminescent behavior upon prolonged laser irradiation, some graphs of how the photoluminescence spectra of our sources varies with time are showed in Fig.~\ref{fig:photoactivation}. In Zhang's article, their study demonstrated that extended exposure to laser light at specific wavelengths induces irreversible modifications in the \ch{C60} molecules, primarily due to photo-induced oxidation processes, leading to a substantial increase in fluorescence intensity and a blue shift in the emission spectrum.

In order to replicate their results, we have measured the spectra of the emitted light of our sample (the one prepared with the 1:100 dilution) when it is shined using large-field imaging with 1.5~mW of power. As the light is collected after the dichroic mirror, only wavelength above 567~nm are recorded in the spectra. Despite this, we can clearly observe how the intensity of the photoluminescence spectrum increases over time, reaching a maximum, and then begins to decrease again until we obtain a spectrum that is practically flat. This latter behavior may be due to a combination of the spectrum shifting towards higher-energy wavelengths, along with the fact that many of the molecules emitting light undergo photobleaching and cease to be emissive, which causes the intensity to drop.

In Fig.~\ref{fig:photoactivation} a), we can see the enhancement of the photoluminescence spectra during the first 150 seconds. We can also see that at the beginning of the measurements we have two peaks, one at 700~nm and the other around 600~nm. Over time, we can observe that the peak initially found around 700~nm shifts toward higher energy wavelengths until it eventually disappears. In Fig.~\ref{fig:photoactivation}~b), we can see the different spectra taken from second 150 until the end of the measurement. Here, we observe how the photoluminescence spectrum begins to decrease almost to the point of vanishing. This process appears to be irreversible, as the shape of the spectra at the end of the measurement is no longer the same as what we observed at the beginning. We can see that the peak initially found around 700~nm never reappears. This suggests that the photoexcitation process is irreversible.

\begin{figure*}[b] 
 \includegraphics[width=1\textwidth]{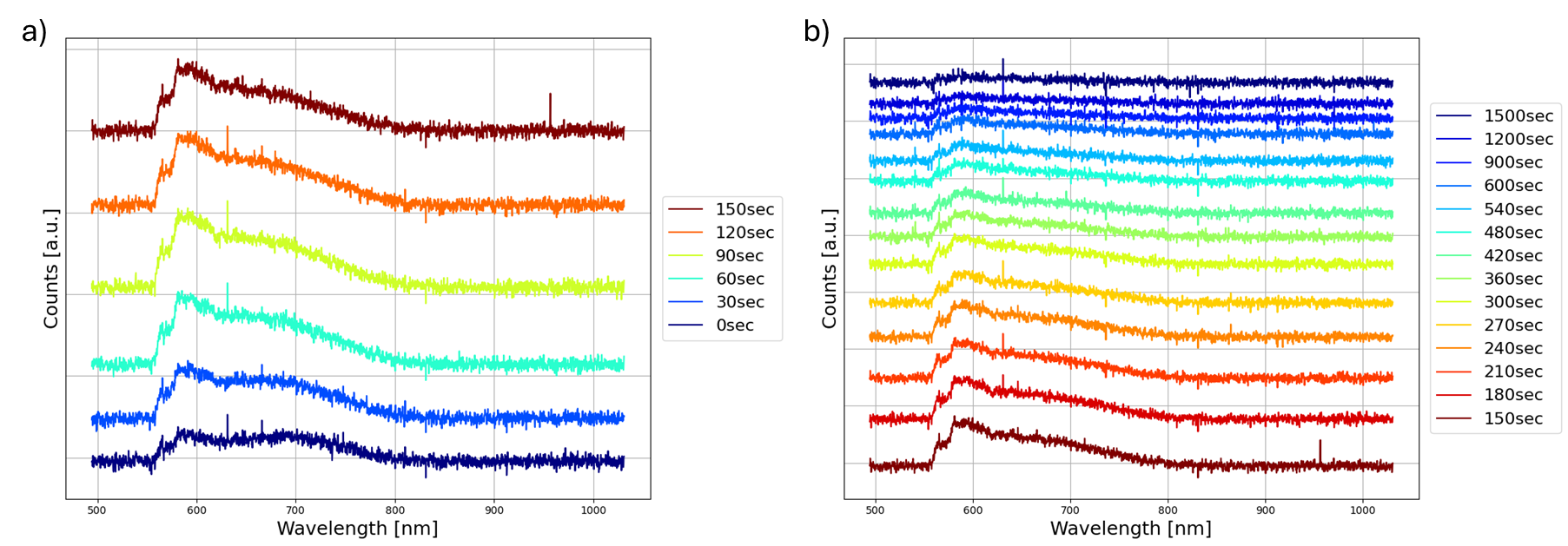}
    \caption{a) Photoluminescence spectra of the sample taken during the first 150~sec. b) Rest of the photoluminescence spectra taken up to a total time of 25~mins. The charts are arranged one on top of the other to aid visualization.}
    \label{fig:photoactivation}
\end{figure*}
If we compare our work with that of Zhang, we notice that in our case, the characteristic waiting times to observe changes in the spectrum are much shorter, on the order of seconds in our work. This may be because we are using a more energetic laser to excite the samples, which could enhance the photoactivation effect, making it occur more rapidly.

\section{Raman spectrometry}
\label{sec:RamanSpectrometry}

Raman spectroscopy measurements were conducted using an excitation wavelength of 532 nm, 
 a 50× objective lens, and a maximum laser power of 6 mW (spectrometer LabRam HR 800, HORIBA). The system provided a spectral resolution of approximately 1 cm$^{-1}$. Acquisition times were varied up to 120 s, with up to 10 accumulations per spectrum. During the measurements using the 532 nm laser, the sample exhibited significant fluorescence background and was prone to photodamage at higher laser powers, therefore low powers of less than 1\% were used to minimize these effects. 
 All the measured peaks in our experiment are summarized in table~\ref{fig:raman_tabla} and shown in Fig.~\ref{fig:raman}.
 
The primary objective of the Raman shift analysis was to confirm that the sample was made of pristine $C_{60}$ and to detect the presence of other carbon structures, such as $C_{70}$ or polymerized \ch{C60}. 

In pristine \ch{C60}, carbon atoms are arranged at the vertices of fused pentagonal and hexagonal rings, forming a highly symmetric, spherical structure~\cite{BETHUNE1990219}. \ch{C60} possesses 174 vibrational modes, of which 10 are Raman-active, corresponding to two $A_g$ modes and eight $H_g$ modes~\cite{Khinevich_2017}. Among these, the $A_g(2)$ mode exhibits the highest Raman intensity, associated with the “pentagonal pinch” vibration. This mode is particularly sensitive to molecular symmetry and environmental changes, with its frequency ranging from approximately 1470~cm$^{-1}$ to 1459~cm$^{-1}$ depending on external factors ~\cite{Khinevich_2017}.  As observed in Fig.~\ref{fig:raman}, the $A_g(2)$ peak appears at 1467~cm$^{-1}$. The presence of a secondary left peak at 1458~cm$^{-1}$, named in this work as $A_g(2)*$, is indicative of polymer formation, likely initiated by exposure to ambient light, as well as the formation of \ch{C60} dimers or $C_{60}O_{2}$ complexes~\cite{DORNERREISEL2022109036, Khinevich_2017}. 

Notably, the $H_g(8)$ frequency measured in our analysis in Fig.~\ref{fig:raman} appears consistent with reported values for $C_{70}$~\cite{BETHUNE1990219,guangyu,Schettino}, suggesting the possible presence of higher fullerene species. The source of smaller peaks highlighted in green in Fig.~\ref{fig:raman} is undetermined, as they don't match with reported $C_{70}$ spectra nor higher forms of \ch{C60}. They could indicate oxidation, degradation of the sample or disorder modes. The broad peak in Fig.~\ref{fig:raman} at 
970~cm$^{-1}$ comes from the measuring system.
The low intensity of the $H_g$ modes and the absence of $H_g(5)$ in Fig.~\ref{fig:raman} could be indicative of the presence of oxidation on the \ch{C60} molecules~\cite{Zygouri}.

Although the samples exhibit minor signs of degradation or oxidation --likely due to air exposure during measurements-- the Raman spectra are consistent with those reported for pristine \ch{C60}, supporting the conclusion that the samples are primarily composed of unmodified fullerene molecules.

\begin{figure*}[t]%
    \centering
    \includegraphics[width=0.7\textwidth]{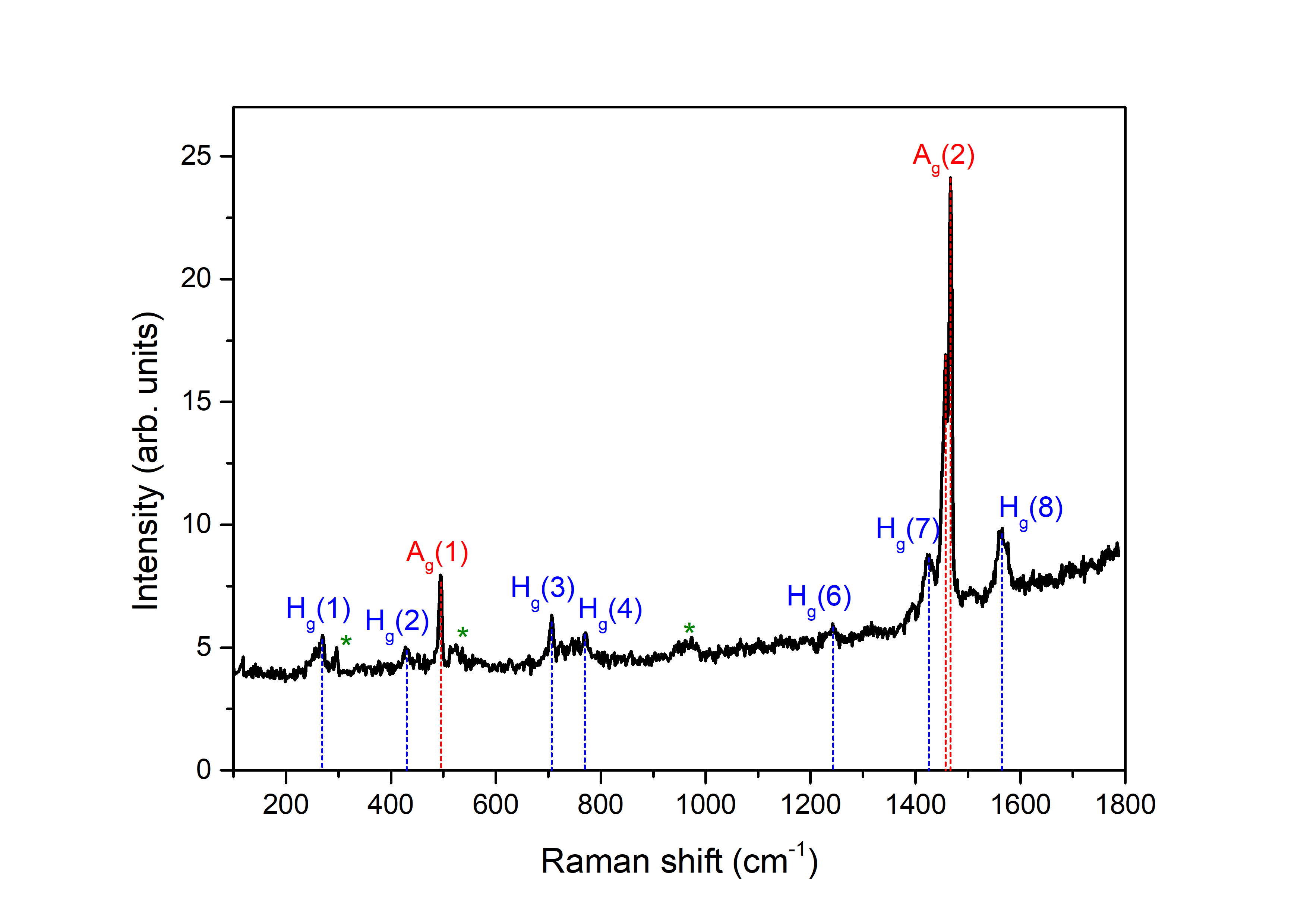}
    \caption{\ch{C60} Raman shift at 1\% of power, acquired at 532 nm wavelength.}%
    \label{fig:raman}%
\end{figure*}

\begin{table}[b]
    \resizebox{\textwidth}{!}{
    \begin{tabular}{|c|c|c|c|c|c|c|c|c|c|c|c|c|c|}
        \hline Symmetry&$A_g(1)$&$A_g(1)*$& $A_g(2)$& $A_g(2)*$& $H_g(1)$& $H_g(2)$& $H_g(3)$& $H_g(4)$& $H_g(5)$&$H_g(6)$& $H_g(7)$& $H_g(8)$\\ \hline 
        532 nm excitation &495&--&1467&1458&269&430&706&770&--&1243&1425&1565\\ \hline
    \end{tabular}}
      \caption{\ch{C60} Raman shift (cm$^{-1}$) at 532~nm excitation wavelength}
    \label{fig:raman_tabla}
\end{table}

\section{Comparison \ch{C60} vs. CQDs}
\label{sec:Comparsion}

To highlight the main differences between colloidal quantum dots (CQDs) and \ch{C60} molecules, we present a comparison of the typical plots of both the second-order autocorrelation function and the emission decay lifetime obtained for each type of single photon source.

For the preparation of the colloidal quantum dot sample, we used a 1:1,000,000 dilution of CdSe/ZnS core-shell colloidal quantum dots (\textit{900219, Sigma-Aldrich}) in dry toluene with 5\% dissolved polystyrene. Once this solution was prepared, a drop was spin-coated onto a gold-coated slide at a speed of 4000~rpm for 1~minute.

\begin{figure*}[t] 
    \centering
    \includegraphics[width=1\textwidth]{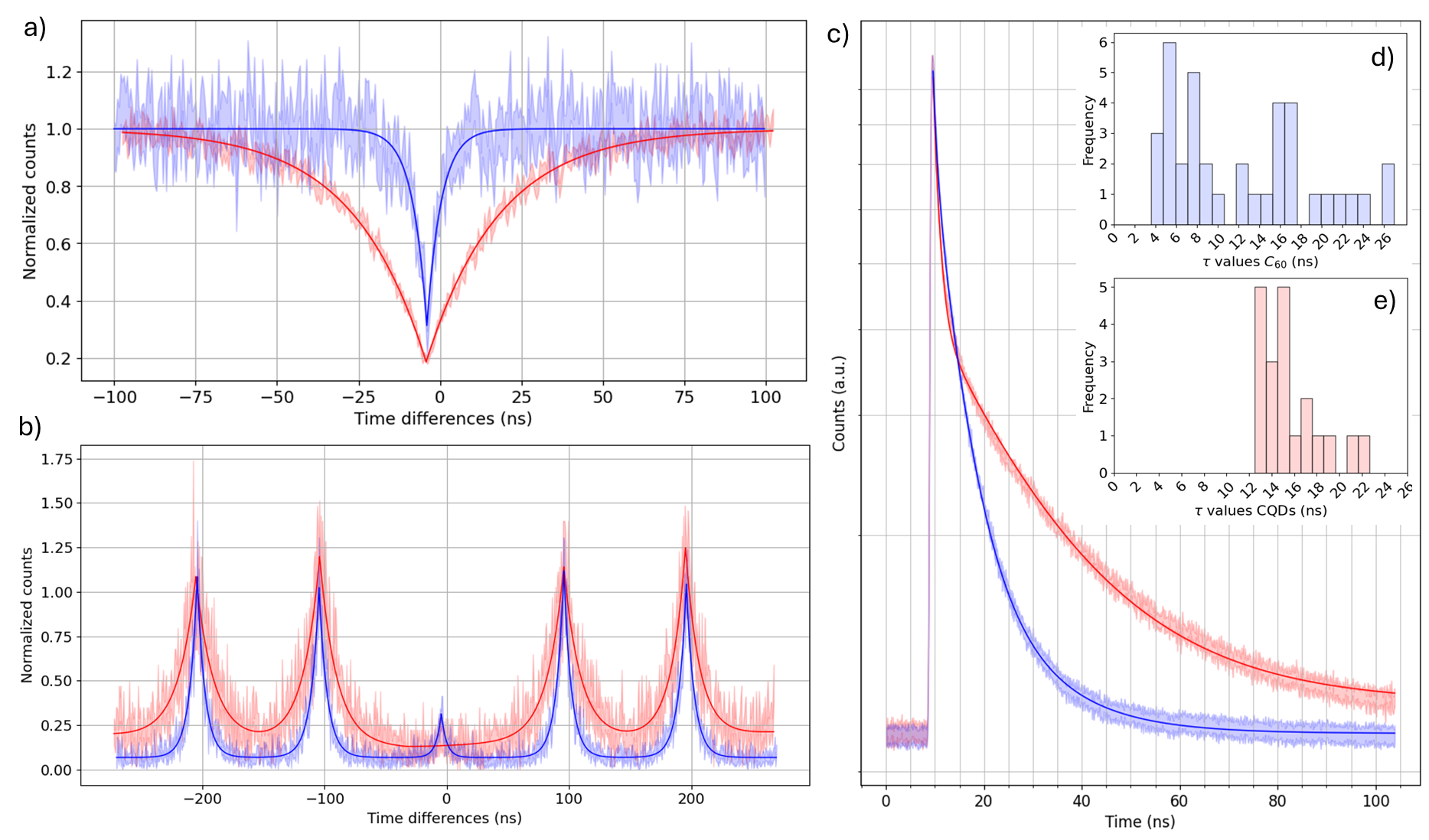}
    \caption{\ch{C60} (blue) vs CQD (red) comparison. a) b) Second-order autocorrelation functions comparison under CW excitation and PW excitation, respectively. In all graphs, the bin width is 500ps. c) Lifetime comparison, where counts on the y-axis are depicted in logarithmic scale. The fit function was adjusted to a bi-exponential decay. d) Histogram of the different decay lifetimes of the \ch{C60} SPS. e) Histogram of the different decay lifetimes of the CdSe/ZnS core-shell colloidal quantum dots. It is worth mentioning that we used the same \ch{C60} single-photon source for all the blue measurements. For the red measurements, the CW second-order autocorrelation function and the decay lifetime  were obtained using the same CQD, while the PW second-order autocorrelation function was measured using a different one.}
    \label{fig:C60_vs_CQD}
\end{figure*}

In Fig.~\ref{fig:C60_vs_CQD} a) we find a comparison between the second-order autocorrelation function under CW excitation using a sample of CQDs (red) and \ch{C60} (blue). We find that the blue graph has a lower decay lifetime, as the dip is more pronounced. Regarding the value of the normalized second-order autocorrelation function and the decay lifetime, using eq.~(1) and eq.~(2), we obtain that, for the CQD, the values are

\begin{equation*}
    g^{(2)}_{norm.}(\tau = \tau_0) = 0.186 \pm 0.007,
    \hspace{0.2cm}\rm{and}\hspace{0.2cm} 
    \tau_X =22.35 \pm 0.37 \hspace{0.1cm}\rm{ns},
\end{equation*}
while for the \ch{C60} the values are
\begin{equation*}
    g^{(2)}_{norm.}(\tau = \tau_0) = 0.312 \pm 0.039,
    \hspace{0.2cm}\rm{and}\hspace{0.2cm} 
    \tau_X = 4.44 \pm 0.37 \hspace{0.1cm}\rm{ns}.
\end{equation*}

On the other hand, for PW excitation, we found the comparison in Fig.~\ref{fig:C60_vs_CQD} b). Now, using the equations~(3) and~(4), we obtain that, for the CQD, the values are

\begin{equation*}
    g^{(2)}_{norm.}(\tau = \tau_0) = 0.304 \pm 0.024,
    \hspace{0.2cm}\rm{and}\hspace{0.2cm} 
    \tau_X = 15.316 \pm 0.736 \hspace{0.1cm}\rm{ns},
\end{equation*}
while for the \ch{C60} the values are
\begin{equation*}
    g^{(2)}_{norm.}(\tau = \tau_0) = 0.304 \pm 0.024,
    \hspace{0.2cm}\rm{and}\hspace{0.2cm} 
    \tau_X = 4.700 \pm 0.692 \hspace{0.1cm}\rm{ns}.
\end{equation*}

For the \ch{C60} (blue) lifetime graph showed in Fig.~\ref{fig:C60_vs_CQD} c) we have fitted the data to a multiexponential function like in the eq.~(5). Obtaining the same results, $\tau_1$ = 0.833~$\pm$~0.015~ns, $\tau_2$ = 4.959~$\pm$~0.095~ns and $\tau_3$ = 13.135~$\pm$~0.450~ns. These results leads us to an averaged decay lifetime of $\tau_{avg.}$ = 4.516~$\pm$~0.079~ns, by using the eq.~(6).

On the other hand, for the CQD (red) lifetime graph showed in Fig.~\ref{fig:C60_vs_CQD} c), we have fitted the values to a biexponential decay~\cite{ihara2019superior}, obtaining $\tau_1$ = 24.164~$\pm$~0.103~ns and $\tau_2$ = 1.334~$\pm$~0.009~ns. Here, $\tau_1$ corresponds to the decay lifetime of the exciton while $\tau_2$ corresponds to the decay lifetime of the biexciton.

Finally, the histograms presented in Fig.~\ref{fig:C60_vs_CQD} d) and e) show the dispersion of the different decay lifetimes observed in both \ch{C60} and CQDs. We can clearly see that \ch{C60}-based SPS have a broad distribution of lifetimes, suggesting different emission environments and greater ease of being disturbed by the environment. While the CQD lifetime distribution is mainly concentrated in the region between 13~ns and 19~ns, the \ch{C60} lifetime distribution ranges from 3~ns to 27~ns.

\end{widetext}
\end{document}